\documentclass[preprintnumbers,nofootinbib,superscriptaddress,twocolumn]{revtex4}
\usepackage[hypertex]{hyperref}
\usepackage[dvipdfmx]{graphicx} 
\usepackage{amsmath,amssymb,amsfonts,dcolumn}

\def\a{\alpha}    \def\d{\delta} \def\D{\Delta}    \def\th{\theta}     \def\m{\mu} \def\n{\nu}      \def\s{\sigma}  \def\t{\tau}       

\def\dg{\dagger}  \def\nn{\nonumber}

\newcommand{\meV}{ {\rm meV} }     

\def\ds{\displaystyle}

  \newcommand{\Lg}{\mathcal{L}}

\newcommand{\Diag}[3]{ \begin{pmatrix} #1 & 0 & 0 \\ 0 & #2 & 0 \\ 0 & 0 & #3 \\\end{pmatrix}}

\begin{document}

\title{Interplay between exact $\m - \t$ reflection symmetries, \\ four-zero texture and universal texture}

\preprint{STUPP-19-239}
\author{Masaki J. S. Yang}
\email{yang@krishna.th.phy.saitama-u.ac.jp}
\affiliation{Department of Physics, Saitama University, 
Shimo-okubo, Sakura-ku, Saitama, 338-8570, Japan}


\begin{abstract} 

In this letter, we consider exact $\mu-\tau$ reflection symmetries for quarks and leptons. 
Fermion mass matrices are assumed to be 
four-zero textures for charged fermions $f = u,d,e$
and a symmetric matrix for neutrinos $\nu_{L}$.
By a bi-maximal transformation,
all the mass matrices lead to $\mu-\tau$ reflection symmetric forms, 
which seperately satisfy $T_{u} \, m_{u,\nu}^{*} \, T_{u} = m_{u,\nu}$   
and $T_{d} \, m_{d,e}^{*} \, T_{d} = m_{d,e}$.

Reconciliation between the $\mu-\tau$ reflection symmetries and observed $\sin \theta_{13}$ 
predicts $\delta_{CP} \simeq \sin^{-1} [\sqrt{m_{e} \over m_{\m}} {c_{13} s_{23} \over s_{13}}] \simeq 203^{\circ}$. 
Moreover, imposition of universal texture $(m_{f})_{11} = 0$ for $f=u,d,\nu,e$  
predicts the normal hierarchy with the lightest neutrino mass $|m_{1}| = 6.26$ or $2.54$ meV.

\end{abstract} 

\maketitle

\section{Introduction}

The T2K experiment recently indicates a nonzero $CP$ violating Dirac phase $\d_{CP}$ \cite{Abe:2014ugx, Abe:2018wpn}.
Then $CP$ violation (CPV) in the lepton sector draws strong attention. 
Among studies of flavor structures, 
$\m-\t$ reflection symmetry \cite{Harrison:2002et,Grimus:2003yn}
is widely studied \cite{Kitabayashi:2005fc, Grimus:2005jk, Farzan:2006vj, Joshipura:2007sf, Adhikary:2009kz, Xing:2010ez, Ge:2010js, Gupta:2011ct, Grimus:2012hu, Joshipura:2015dsa, Xing:2015fdg, He:2015afa, Chen:2015siy, Samanta:2017kce, Xing:2017cwb, Nishi:2018vlz, Nath:2018hjx, Sinha:2018xof}
because it predicts bi-maximal mixing $\th_{23} = 45^{\circ}$ and the maximal CPV Dirac phase $\d_{CP} = \pm \pi/2$.

In this context, universal texture \cite{Koide:2002cj,Koide:2003rx,Joshipura:2005vy,Joshipura:2009tg}
that imposes $\m-\t$ permutation symmetry or $2-3$ symmetry for all the SM fermions 
is appealing. 
However, universal  $\m-\t$ reflection symmetry naively fails because it predicts zero $CP$ phase $\d_{CP} = 0$. 


In this letter, we consider exact $\m-\t$ reflection symmetries for quarks and leptons. 
By a bi-maximal transformation,
four-zero textures \cite{Fritzsch:1995nx, Xing:1996hi, Kang:1997uv, Mondragon:1998gy, Nishiura:1999yt, Matsuda:1999yx, Fritzsch:1999ee, Fritzsch:2002ga, Xing:2003zd, Xing:2003yj, Bando:2004hi, Xing:2015sva} lead to a $\m - \t$ reflection symmetric form in a particular basis.
In this basis, up- and down-type quark mass matrices from four-zero textures 
separately satisfy exact $\m - \t$ reflection symmetries, $T_{u,d} \, m_{u,d}^{*} \, T_{u,d} = m_{u,d}$.  
These symmetries realize the maximal $CP$ violation in the Fritzsch -- Xing parameterization \cite{Fritzsch:1997fw}. 

The same symmetries also hold in the lepton sector, $T_{u,d} \, m_{\n,e}^{*} \, T_{u,d} = m_{\n,e}$.
In order to reconcile the $\m - \t$ reflection symmetries and observed $\sin \th_{13}$, 
the maximal $CP$ violation is discarded and it predicts 
$\d_{CP} \simeq \sin^{-1} [\sqrt{m_{e} \over m_{\m}} {c_{13} s_{23} \over s_{13}}] \simeq 203^{\circ}$. 
This value is rather close to the best fit for the normal hierarchy and in the $1 \s$ region $\d_{CP} / ^{\circ} = 217^{+40}_{-28}$ \cite{Esteban:2018azc}.
This result does not depend on the mass ordering of neutrinos. 

Moreover, assuming universal texture $(m_{f})_{11} = 0$ for $f=u,d,\n,e$ 
and small 2-3 mixing of the lepton mass matrix, we obtain the lightest neutrino mass $|m_{1}| = 6.26$ or $2.54$ meV. 
Each value corresponds to the Majorana phases $|\a_{3} - \a_{2}| = 0$ or $\pi$, 
and the result only holds in the case of the normal hierarchy (NH).
Besides, for the inverted hierarchical case, the solutions do not have real values and then contradict with the reflection symmetries.


This paper is organized as follows. 
In the next section, we review four-zero textures and $\m-\t$ reflection symmetry. 
In Sec.~3, $\m-\t$ symmetries are imposed on the lepton sector. 
The final section is devoted to conclusions.

\section{Four-zero texture and $\m - \t$ reflection symmetry }

In this section, we review four-zero textures 
and its interplay of the $\m-\t$ reflection symmetry \cite{Harrison:2002et,Grimus:2003yn}. 
First of all, the phenomenological mass matrices of the SM fermions $f= u,d,e$ and neutrinos $\n_{L}$ are defined by 
\begin{align}
\Lg \ni  \sum_{f} -  \bar f_{Li } m_{f ij} f_{Rj} - \bar \n_{L i} m_{\n ij} \n_{L j}^{c} + {\rm h.c.} \, .
\end{align}
Diagonalization of the mass matrices $m_{f} = U_{Lf} m_{f}^{\rm diag} U_{Rf}^{\dg}$ 
leads to the following CKM and  MNS mixing matrices 
\begin{align}
V_{\rm CKM}  = U_{L u }^{\dg} U_{Ld} , ~~~ 
U_{\rm MNS} = U_{Le}^{\dg} U_{L \n} .  
\end{align}

Since both the matrices have a large Dirac phase, 
the maximal $CP$ violation have been discussed \cite{Fritzsch:1985yv, Fritzsch:1995nx}. 
$CP$ phases depend on the basis of the fermions. 
In particular, the phase of the CKM matrix 
becomes almost maximal in the Fritzsch--Xing parameterization \cite{Fritzsch:1997fw}:
\begin{align}
V_{\rm CKM} &= 
\begin{pmatrix}
c_{u} & s_{u} & 0 \\
- s_{u} & c_{u} & 0 \\
 0 & 0 & 1
\end{pmatrix}
\begin{pmatrix}
e^{- i \d_{\rm FZ}} & 0 & 0 \\
 0 &  c_{q} & s_{q} \\
 0 & - s_{q} & c_{q} \\
\end{pmatrix}
\begin{pmatrix}
c_{d} & -s_{d} & 0 \\
s_{d} & c_{d} & 0 \\
 0 & 0 & 1
\end{pmatrix}  ,
\label{FZCKM}
\end{align}
where $c_{f} \equiv \cos \th_{f}, s_{f} \equiv \sin \th_{f}.$
The best fit values are calculated by the way of Ref.~\cite{Parodi:1999nr}
\begin{align}
s_{u} = 0.0863 , ~~ s_{d} = 0.212, ~~ s_{q} = 0.0423, ~~ \d_{\rm FZ} = 87.9^{\circ} . 
\end{align}

Although the original Kobayashi--Maskawa parameterization \cite{Kobayashi:1973fv}
has a similar result $\d_{\rm KM} \simeq \pi/ 2$ \cite{Koide:2004gj},
the Fritzsch -- Xing parameterization has a reasonable physical view because it 
factorizes the large mixing in 1-2 generations 
and the small one in 2-3 generations.

If we assume $m_{u,d}$ are Hermitian matrices, 
Eq.~(\ref{FZCKM}) strongly suggests that 
the mass matrices of quarks have ``four-zero texture'' or ``modified Fritzsch texture'' 
\cite{Fritzsch:1995nx}, 
\begin{align}
m_{u} &= 
\Diag{i}{1}{1}
\begin{pmatrix}
0 & C_{u} & 0 \\ 
 C_{u} & \tilde B_{u} &  B_{u} \\
0 & B_{u} & A_{u} 
\end{pmatrix}
\Diag{-i}{1}{1} , \label{mu0} \\ 
m_{d} & =  
\begin{pmatrix}
0 & C_{d} & 0 \\ 
C_{d} & \tilde B_{d} & B_{d} \\
0 & B_{d}  & A_{d} 
\end{pmatrix} , \label{md0}
\end{align}
with real parameters $A_{f} > B_{f} > \tilde B_{f} \gg C_{f}$. 
For the later convenience, 
the relative phase is pressed on $m_{u}$. 
The forms of mass matrices~(\ref{mu0}) and (\ref{md0}) are almost consistent with the
latest full parameter scan of four-zero textures \cite{Xing:2003yj, Xing:2015sva}.
In this case, the rotation matrices $V_{u,d}$ at leading order is written by the mass eigenvalues
and  parameters $r_{u,d} \equiv A_{u,d} / m_{t,b}$ \cite{Xing:2015sva}; 
\begin{align}
V_{u} & \simeq 
\begin{pmatrix}
 1 & 0 & 0 \\
 0 & \sqrt{r_{u}} & \sqrt{1-r_{u}} \\[3pt]
 0 & - \sqrt{1-r_{u}} & \sqrt{r_{u}} \\
\end{pmatrix}
\begin{pmatrix}
 1 & \ds - \sqrt{m_{u} \over m_{c}} & 0 \\
\ds \sqrt{m_{u} \over m_{c}} & 1 & 0 \\
 0 & 0 & 1 \\
\end{pmatrix} ,
~~~ \\
V_{d} & \simeq 
\begin{pmatrix}
 1 & 0 & 0 \\
 0 & \sqrt{r_{d}} & \sqrt{1-r_{d}} \\[3pt]
 0 & - \sqrt{1-r_{d}} & \sqrt{r_{d}} \\
\end{pmatrix}
\begin{pmatrix}
 1 & \ds - \sqrt{m_{d} \over m_{s}} & 0 \\
 \ds \sqrt{m_{d} \over m_{s}} & 1 & 0 \\
 0 & 0 & 1 \\
\end{pmatrix} .
\end{align}
Then, an approximate form of the CKM matrix $V_{\rm CKM} = U_{u}^{\dg} U_{d}$ is found to be
\begin{align}
& V_{\rm CKM}  = V_{u}^{T} \Diag{-i}{1}{1} V_{d}. 
\label{VCKM2}
\end{align}
In the Fritzsch--Xing parameterization~(\ref{FZCKM}), one directly obtains
\begin{align}
s_{u} &\simeq \sqrt{m_{u} \over m_{c}} , ~  s_{d} \simeq \sqrt{m_{d} \over m_{s}} , \\
 s_{q} &\simeq \sqrt{r_{u} (1-r_{d})} - \sqrt{r_{d} (1-r_{u})} , ~ \d_{\rm FZ}  \fallingdotseq \pi/2. 
\end{align}
It predicts $V_{cb}$ and $V_{ts}$ at leading order as follows
\begin{align}
|V_{cb}| \simeq |V_{ts}| \simeq s_{q}  .
\end{align}

A bi-maximal transformation of the mass matrices by the following $U_{BM}$, 
\begin{align}
m_{f}^{BM} \equiv  U_{BM}^{\dag} m_{f} U_{BM}, 
~~~
U_{BM} \equiv
\begin{pmatrix}
 1 & 0 & 0 \\
 0 & \frac{i}{\sqrt{2}} & \frac{i}{\sqrt{2}} \\
 0 & -\frac{1}{\sqrt{2}} & \frac{1}{\sqrt{2}} \\
\end{pmatrix} , 
\label{bimaximal}
\end{align}
leads to 
\begin{align}
m_{u}^{BM} & = 
\begin{pmatrix}
 0 & -\frac{{C_{u}}}{\sqrt{2}} & -\frac{{C_{u}}}{\sqrt{2}} \\
- \frac{{C_{u}}}{\sqrt{2}} & \frac{{\tilde B_{u}}}{2}+\frac{{A_{u}}}{2} & \frac{{\tilde B_{u}}}{2}-\frac{{A_{u}}}{2}-i {B_{u}} \\
- \frac{{C_{u}}}{\sqrt{2}} &\frac{{\tilde B_{u}}}{2}-\frac{{A_{u}}}{2} +  i {B_{u}} & \frac{{\tilde B_{u}}}{2}+\frac{{A_{u}}}{2} \\
\end{pmatrix} ,  \label{mup}
\\
m_{d}^{BM} & = 
\begin{pmatrix}
 0 & \frac{i {C_{d}}}{\sqrt{2}} & \frac{i {C_{d}}}{\sqrt{2}} \\
 -\frac{i {C_{d}}}{\sqrt{2}} & \frac{{\tilde B_{d}}}{2}+\frac{{A_{d}}}{2} &\frac{{\tilde B_{d}}}{2}-\frac{{A_{d}}}{2}  -i  {B_{d}}\\
 -\frac{i {C_{d}}}{\sqrt{2}} & \frac{{\tilde B_{d}}}{2}-\frac{{A_{d}}}{2} + i  {B_{d}} & \frac{{\tilde B_{d}}}{2}+\frac{{A_{d}}}{2} \\
\end{pmatrix} . \label{mdown}
\end{align}
These matrices~(\ref{mup}), (\ref{mdown}) separately satisfy  exact $\m - \t$ reflection symmetries:
\begin{align}
T_{u} (m_{u}^{BM})^{*} T_{u} = m_{u}^{BM}  , ~~~ 
T_{d} (m_{d}^{BM})^{*} T_{d} = m_{d}^{BM}  , ~~~ 
\label{mutausym}
\end{align} 
where 
\begin{align}
T_{u}= 
\begin{pmatrix}
 1 & 0 & 0 \\
 0 & 0 & 1 \\
 0 & 1 & 0 \\
\end{pmatrix} , ~~~
T_{d} =
\begin{pmatrix}
 1 & 0 & 0 \\
 0 & 0 & -1 \\
 0 & -1 & 0 \\
\end{pmatrix} .
\end{align}
The form of matrices~(\ref{mup}), (\ref{mdown})
has been indicated a study of universal texture \cite{Koide:2002cj}.
However, they considered no symmetry and  these forms were treated as a phenomenological description. 

\section{$\m- \t$ reflection symmetry in the lepton sector }

In this section, 
we impose the symmetries~(\ref{mutausym}) 
on the lepton sector and research some predictions. 
If the MNS matrix has the same form to the CKM matrix~(\ref{VCKM2}), 
\begin{align}
U_{0} = 
V_{e}^{T}
 \Diag{+i}{1}{1} 
V_{\n0} , 
\label{U0}
\end{align}
where
\begin{align}
V_{e} & \simeq
\begin{pmatrix}
1 & 0 & 0 \\
0 & \sqrt{r_{e}} & \sqrt{1-r_{e}} \\
0 & \ds - \sqrt{1-r_{e}} & \sqrt{r_{e}}
\end{pmatrix}
\begin{pmatrix}
1 & \ds - \sqrt{m_{e} \over m_{\m}} & 0 \\
\ds \sqrt{m_{e} \over m_{\m}} & 1 & 0 \\
0 & 0 & 1
\end{pmatrix} , \label{Vre} \\  
V_{\n 0} & \simeq
\begin{pmatrix}
 1 & 0 & 0 \\
 0 & c_{23} &  s_{23} \\
 0 & - s_{23} & c_{23} \\
\end{pmatrix}
\begin{pmatrix}
c_{12} & - s_{12} & 0 \\
  s_{12} & c_{12} & 0 \\
 0 & 0 & 1 \\
\end{pmatrix} ,
\label{Vn}
\end{align}
with $r_{e} \equiv A_{e}/m_{\tau}$ in Eq.~(\ref{nosym}). 
It predicts the maximal $CP$ violation $ \d^{PDG} \fallingdotseq - \pi/2$ and
\begin{align}
 s_{23}^{PDG} \simeq 1/ \sqrt 2 , ~~~ s_{13}^{PDG} \simeq s_{23}^{PDG} \sqrt{m_{e} / m_{\m}} \simeq 0.05 ,
\end{align}
in the standard PDG parameterization. 
However, the small $\sin \th_{13}$ disagrees with the observation.
They are well known results in the universal texture \cite{Koide:2002cj, Koide:2003rx}.

In order to derive a proper $\sin \th_{13}$, 
we change the mixing matrix Eqs.~(\ref{U0}) and~(\ref{Vn}) in the following way. 
\begin{align}
U &= 
V_{e}^{T}
 \Diag{-i}{1}{1} 
V_{\n}, \label{UVn0} \\
~~~ 
V_{\n} &= 
\begin{pmatrix}
 1 & 0 & 0 \\
 0 & c_{23} & s_{23} \\
 0 & - s_{23} & c_{23} \\
\end{pmatrix}
\begin{pmatrix}
c_{13} & 0 & s_{13} \\
 0 & 1 & 0 \\
- s_{13} & 0 & c_{13} \\
\end{pmatrix}
\begin{pmatrix}
c_{12} & s_{12} & 0 \\
 - s_{12} & c_{12} & 0 \\
 0 & 0 & 1 \\
\end{pmatrix} .
\label{UVn}
\end{align}
In Eqs.~(\ref{UVn0}) and (\ref{UVn}), the sign of the phase and $s_{12}$ are changed in the same way to the PDG parameterization 
(the sign of $J^{PDG}$~(\ref{JPDG}) is not changed).  
The Majorana phases are omitted here and discussed later. 
The 2-3 mixing of the $V_{e}$
can be absorbed to that of the $V_{\n}$. 
Then, 
\begin{align}
U & = 
\begin{pmatrix}
1 &  \ds \sqrt{m_{e} \over m_{\m}} & 0 \\
- \ds \sqrt{m_{e} \over m_{\m}} & 1 & 0 \\
0 & 0 & 1
\end{pmatrix} \nn \\ & \times 
\begin{pmatrix}
-i & 0 & 0 \\
 0 & c_{23} & s_{23} \\
 0 & - s_{23} & c_{23} \\
\end{pmatrix}
\begin{pmatrix}
c_{13} & 0 & s_{13} \\
 0 & 1 & 0 \\
- s_{13} & 0 & c_{13} \\
\end{pmatrix}
\begin{pmatrix}
c_{12} & s_{12} & 0 \\
 - s_{12} & c_{12} & 0 \\
 0 & 0 & 1 \\
\end{pmatrix} .
\label{MNSmatrix}
\end{align}
It leads to 
\begin{align}
| U_{e3} | = | s_{13}^{PDG} | \simeq |  i s_{13} - \sqrt{m_{e} \over m_{\m}} c_{13} s_{23} | .
\end{align}
From the parameter $\sqrt{m_{e} / m_{\m}} \simeq 0.07$ and 
the following values from the latest global fit \cite{Esteban:2018azc},
\begin{align}
\th_{23}^{PDG} = 49.7^{\circ} , ~~~~
\th_{12}^{PDG} = 33.82^{\circ} , ~~~~
\th_{13}^{PDG} = 8.61^{\circ} ,
\end{align}
one obtains $c_{13} \simeq c_{13}^{PDG}$, $s_{23} \simeq s_{23}^{PDG}$
and 
\begin{align}
s_{13} = \pm \sqrt{ (s_{13}^{PDG})^{2} - {m_{e} \over m_{\m}} (c_{13}^{PDG})^{2} (s_{23}^{PDG})^{2} } = \pm 0.140.  \label{eq22}
\end{align}
The sign $\pm$ in Eq.~(\ref{eq22}) corresponds to the sign of $\cos \d_{CP}$. 
We adopt $s_{13} = -0.140$ because the latest global fit found $\cos \d_{CP} < 0$ \cite{Esteban:2018azc}. 

The absolute values of elements in reconstructed $U_{\rm MNS}$ 
are found to be 
\begin{align}
|U_{\rm MNS}| = 
\begin{pmatrix}
 0.821 & 0.550 & 0.149 \\
 0.275 & 0.596 & 0.754 \\
 0.500 & 0.584 & 0.640 \\
\end{pmatrix}.
\end{align}
All of the components is in the range of 3 $\s$.

On the analogy of quark masses, the lepton mass matrices $m_{\n, e}$ 
which predict the mixing matrix~(\ref{UVn}) will be the following forms
\begin{align}
m_{\n} & = 
\Diag{-i}{1}{1}
\begin{pmatrix}
a_{\n} & b_{\n} & c_{\n} \\ 
b_{\n} & d_{\n} & e_{\n} \\
c_{\n} & e_{\n}  & f_{\n} 
\end{pmatrix}
\Diag{-i}{1}{1} \label{nosym2}
, \\
m_{e} & =
\begin{pmatrix}
0 & C_{e} & 0 \\ 
C_{e} &\tilde B_{e} & B_{e} \\
0 &  B_{e}  & A_{e} 
\end{pmatrix} , \label{nosym}
\end{align}
with real parameters $a_{\n} \sim f_{\n}$. 
These matrices have no symmetry 
at first glance. 
However, 
the bi-maximal transformation in Eq.~(\ref{bimaximal})
\begin{align}
m_{\n}^{BM} &\equiv 
U_{BM}^{\dg} m_{\n} U_{BM}^{*} , ~~~
m_{e}^{BM} \equiv  U_{BM}^{\dg} m_{e}  U_{BM} ,
\end{align}
 leads to 
\begin{align}
m_{\n}^{BM} &
= 
\begin{pmatrix}
 -a_{\n} & {1\over \sqrt{2}} (b_{\n} - i c_{\n}) &  {1\over \sqrt{2}} (b_{\n} + i c_{\n})  \\
 {1\over \sqrt{2}} (b_{\n} - i c_{\n})  & \frac{f_{\n}}{2} -\frac{d_{\n}}{2}+ i e_{\n} & -\frac{f_{\n}}{2}-\frac{d_{\n}}{2} \\
 {1\over \sqrt{2}} (b_{\n} + i c_{\n})  & -\frac{f_{\n}}{2}-\frac{d_{\n}}{2} & \frac{f_{\n}}{2} -\frac{d_{\n}}{2}- i e_{\n}\\
\end{pmatrix} \label{mnu}
, \\
m_{e}^{BM} &=
\begin{pmatrix}
 0 & \frac{i {C_{e}}}{\sqrt{2}} & \frac{i {C_{e}}}{\sqrt{2}} \\
 -\frac{i {C_{e}}}{\sqrt{2}} & \frac{{\tilde B_{e}}}{2}+\frac{{A_{e}}}{2} &\frac{{\tilde B_{e}}}{2}-\frac{{A_{e}}}{2}  -i  {B_{e}}\\
 -\frac{i {C_{e}}}{\sqrt{2}} & \frac{{\tilde B_{e}}}{2}-\frac{{A_{e}}}{2} + i {B_{e}} & \frac{{\tilde B_{e}}}{2}+\frac{{A_{e}}}{2} \\
\end{pmatrix} . \label{me}
\end{align}
Eqs.~(\ref{mnu}) and (\ref{me}) also separately satisfy  exact $\m - \t$ reflection symmetries~(\ref{mutausym}):
\begin{align}
T_{u} (m_{\n}^{BM})^{*} T_{u} = m_{\n}^{BM}  , ~~~ 
T_{d} (m_{e}^{BM})^{*} T_{d} = m_{e}^{BM}  .
\label{refsym}
\end{align}
Therefore, in this basis, all of quarks and leptons satisfy 
the exact $\m-\t$ reflection symmetries. 
Note that the $\m-\t$ symmetry is not imposed on $m_{\n}$
in the basis of four-zero texture~(\ref{nosym2}).

\subsection{Dirac phase $\d_{CP}$}

In order to show the Dirac phase $\d_{CP}$, 
we evaluate the Jarskog invariant \cite{Jarlskog:1985ht}, 
\begin{align}
J^{PDG} = 
\sin \d_{CP} s_{12}^{PDG} c_{12}^{PDG} s_{13}^{PDG} (c_{13}^{PDG})^{2} s_{23}^{PDG}  c_{23}^{PDG}.
\label{JPDG}
\end{align}
Since the phase $\d_{CP}$ vanishes in a limit of $ \sqrt{m_{e} / m_{\m}} \to 0$, 
 the invariant should be proportional to the parameter. 
The invariant is evaluated from Eq.~(\ref{MNSmatrix}) as
\begin{align}
J &= - {\rm Im} \,  [U_{\m 3} U_{\t 2} U_{\m 2}^{*} U_{\t 3}^{*}] \\
& \simeq  
\sqrt{m_{e} / m_{\m}} \,  c_{13} c_{23}  [ - c_{12} s_{12} s_{23}^2 + s_{13} c_{23}  s_{23} (c_{12}^2  - s_{12}^2 ) \nn \\
 & + s_{13}^2 c_{12} s_{12} c_{23}^2 ] = - 0.0130 ,  \\
 & \simeq - \sqrt{m_{e} / m_{\m}} \, c_{13} c_{23} c_{12} s_{12} s_{23}^2 = (-0.0120) . 
 \label{exp}
\end{align}
The value with (without) parentheses is induced from full calculation (only leading order). 
Since $s_{13} = - 0.14$ is small, 
$J^{PDG}$~(\ref{JPDG}) and the perturbative expansion~(\ref{exp}) gives an approximate formula of $\sin \d_{CP}$,  
\begin{align}
\sin \d_{CP} &\simeq \sqrt{m_{e} \over m_{\m} } {c_{13} s_{23} \over s_{13}}
\simeq -0.390 ~ (-0.360) .
\end{align}
cos $\d_{CP}$ is obtained as 
\begin{align}
& \cos \d_{CP} =  { |U_{22}^{PDG}|^{2} - (s_{12}^{PDG} s_{13}^{PDG} s_{23}^{PDG})^{2} - (c_{12}^{PDG} c_{23}^{PDG})^{2} \over - 2 s_{12}^{PDG} s_{13}^{PDG} s_{23}^{PDG} c_{12}^{PDG} c_{23}^{PDG} }    \\
&= {|U_{22}|^{2} (1 - |U_{13}|^{2})^{2} - |U_{13}|^{2} |U_{12}|^{2} |U_{23}|^{2} - |U_{11}|^{2} |U_{33}|^{2} \over - 2 |U_{13}| |U_{12}| |U_{23}| |U_{11}| |U_{33}|} \\ & = -0.920. 
\end{align}
Therefore
\begin{align}
\d_{CP} &\simeq 203^{\circ} ~ (201^{\circ}) .
\end{align}
This value is rather close to the best fit for the normal hierarchy and in the $1 \s$ region $\d_{CP} / ^{\circ} = 217^{+40}_{-28}$ \cite{Esteban:2018azc}.
This result does not depend on the mass ordering of neutrinos. 

\subsection{Majorana phases, universal zero texture and masses} 

The standard PDG convention of the Majorana phases is
\begin{align}
U_{\rm MNS} = U P ,  ~~ P \equiv  {\rm diag} (1 ,  e^{ i \a_{2} / 2} , e^{ i \a_{3} / 2}) . 
\end{align}
The neutrino mass matrix $m_{\n}$ reconstructed in the four-zero basis 
is obtained as 
\begin{align}
m_{\n} = \Diag{-i}{1}{1} V_{\n} P \Diag{m_{1}}{m_{2}}{m_{3}} P V_{\n}^{T} \Diag{-i}{1}{1} . 
\label{fullmn}
\end{align}
If this mass matrix with Majorana phases
satisfies the symmetry Eq.~(\ref{mutausym}), 
 $\a_{2,3} /2 = n \pi /2$ $(n=0,1,2, ...)$ should be hold. 
This result agrees to the previous studies by Xing {\it et.\,al.} \cite{Xing:2017cwb, Nath:2018hjx}. 

Note that the 2-3 mixing of $V_{\n}$ and $V_{e}$ cannot be determined independently.
 We assume that of $V_{e}$ is small (equivalently, $\sqrt{1-r_{e}} \simeq m_{\m} / m_{\t} $ in Eq.~(\ref{Vre}) ).
Moreover, imposition of  universal texture $(m_{f})_{11} = 0$ for $f = u,d,\n ,e$ \cite{Koide:2002cj} determines the mass eigenvalues $m_{1,2,3}$ from a condition
\begin{align}
m_{1} = {-e^{i \alpha_{2}} m_{2} s_{12}^{2} - e^{i \alpha_{3}} m_{3} t_{13}^{2} \over c_{12}^2  } ,
\label{m1}
\end{align}
where $t_{13} \equiv \tan \th_{13}$. 
For the normal hierarchical case, 
\begin{align}
|m_{1}| &= 6.20 \, \meV ~~  {\rm for} ~~  (\a_{2}, \a_{3}) = (0,0)  \text{ or } (\pi, \pi ) \\
&= 2.54 \, \meV ~~ {\rm for} ~~ (\a_{2}, \a_{3}) = (0, \pi) \text{ or } (\pi, 0)  .
\end{align}
 Here, we used the mass differences from the global fit \cite{Esteban:2018azc}
\begin{align}
\D m_{21}^{2} = 73.9 \, [\meV^{2}], ~~~ 
\D m_{31}^{2} =  2525 \, [\meV^{2}]. 
\end{align}
Besides, for the inverted hierarchical case, the solutions of
 Eq.~(\ref{m1}) do not have real values and then contradict with the reflection symmetries.

Finally, the effective mass ${m_{ee}}$ of the double beta decay is obtained as \cite{Vergados:1985pq}
\begin{align}
|m_{ee}| &= \left| \sum_{i=1}^{3} m_{i} U_{ei}^{2} \right| \\
& = |(c_{13}^{PDG})^{2} [m_{1} (c_{12}^{PDG})^{2} + m_{2} (s_{12}^{PDG})^{2} e^{i \a_{2}}] \nn \\ 
&~~~ + m_{3} (s_{13}^{PDG})^{2} e^{i (\a_{3} -2 \d)}  | , \\
&= 0.17 \meV  ~~ {\rm for} ~~ (\a_{2}, \a_{3}) = (0,0)  \text{ or } (\pi, \pi ) , \\
 &= 1.24 \meV  ~~ {\rm for} ~~ (\a_{2}, \a_{3}) = (0, \pi) \text{ or } (\pi, 0)  . 
\end{align}
These values are rather small than other models because 
it vanishes in a limit of $(m_{\n})_{11} = \sqrt {m_{e} / m_{\m}} = 0$. 
In particular, 
the phase factor $-i$ in Eq.~(\ref{fullmn}) 
generates destructive interferences for $\a_{2} = \a_{3}$.

\section{Conclusions} 

In this letter, we consider exact $\m-\t$ reflection symmetries for quarks and leptons. 
By a bi-maximal transformation,
up- and down-type quark mass matrices with 
four-zero textures separately satisfy exact $\m - \t$ reflection symmetries, $T_{u,d} \, m_{u,d}^{*} \, T_{u,d} = m_{u,d}$. 

The same symmetries also hold in the lepton sector, $T_{u,d} \, m_{\n,e}^{*} \, T_{u,d} = m_{\n,e}$. 
Reconciliation between the $\m - \t$ reflection symmetries and observed $\sin \th_{13}$ 
predicts $\d_{CP} \simeq \sin^{-1} [\sqrt{m_{e} \over m_{\m}} {c_{13} s_{23} \over s_{13}}] \simeq 203^{\circ}$. 
This value is rather close to the best fit for the normal hierarchy and in the $1 \s$ region $\d_{CP} / ^{\circ} = 217^{+40}_{-28}$.

Moreover, assuming universal texture $(m_{f})_{11} = 0$ for $f=u,d,\n,e$ 
and small 2-3 mixing of the lepton mass matrix, we obtain the lightest neutrino mass $|m_{1}| = 6.26$ or $2.54$ meV. 
This result only holds in the case of the normal hierarchy, 
because the solutions  contradict with the reflection symmetries for the inverted hierarchical case.

\section*{Acknowledgement}

This study is financially supported by the Iwanami Fujukai Foundation,  and 
JSPS KAKENHI Grants No. 20K14459.


\end{document}